\pdfoutput=1
\documentclass[pra,twocolumn,aps,10pt]{revtex4-1}
\usepackage{amsmath}
\usepackage{amsfonts}
\usepackage{amssymb}
\usepackage{graphicx}
\usepackage{hyperref}
\begin{document}

%========= TITLE =====================================================

\title{Absolute L-shell ionization and X-ray production cross sections
of Lead and Thorium by 16-45 keV electron impact}

%==================== AUTHORS ================================%
%==================== AUTHORS ================================%

\author{H. V. Rahangdale}
\email{hitesh.rahangdale@saha.ac.in} 
\altaffiliation{ Applied Nuclear Physics Division, Saha Institute of 
Nuclear Physics, 1/AF, Bidhannagar, Kolkata-700064, India}
\author{P. K. Das}
%\email{pradiptak.das@saha.ac.in} 
\altaffiliation{ Applied Nuclear Physics Division, Saha Institute of 
Nuclear Physics, 1/AF, Bidhannagar, Kolkata-700064, India}
\author{S. De}
%\email{sankar.de@saha.ac.in} 
\altaffiliation{ Applied Nuclear Physics Division, Saha Institute of 
Nuclear Physics, 1/AF, Bidhannagar, Kolkata-700064, India}
\author{J. P. Santos}
%\email{jps@fct.unl.pt}

\altaffiliation{ Laborat\'{o}rio de Instrumenta\c{c}\~{a}o, Engenharia
Biom\'{e}dica eF\'{\i}sica da Radia\c{c}\~{a}o (LIBPhys-UNL), Departamento de F\'{\i}sica, Faculdade de
Ci\^{e}ncias e Tecnologia, FCT, Universidade Nova de Lisboa, 2829-516
Caparica, Portugal}

% \altaffiliation{ Centro de F\'isica At\'omica, CFA, Departamento de  
% F\'isica, Faculdade de Ci\^encias e Tecnologia, FCT, Universidade 
% Nova de Lisboa, 2829-516 Caparica, Portugal}
\author{D. Mitra}
%\email{debasis.mitra51@gmail.com}
\altaffiliation{ Department of Physics, University of Kalyani, 
Kalyani, Nadia-741235, India}
\author{M. Guerra}
%\email{mguerra@campus.fct.unl.pt} 
\altaffiliation{ Centro de F\'isica At\'omica, CFA, Departamento de  
F\'isica, Faculdade de Ci\^encias e Tecnologia, FCT, Universidade 
Nova de Lisboa, 2829-516 Caparica, Portugal}
\author{S. Saha}
%\email{satyajit.saha@saha.ac.in} 
\altaffiliation{ Applied Nuclear Physics Division, Saha Institute of 
Nuclear Physics, 1/AF, Bidhannagar, Kolkata-700064, India}

%==================== ABSTRACT==========================%
%==================== ABSTRACT==========================%

\begin{abstract}

The absolute L subshell specific electron impact ionization cross
sections near the ionization threshold ($16 < E < 45$~keV) of Lead and
Thorium are obtained from the measured L X-ray production cross
sections. Monte Carlo simulation is done to account for the effect of
the backscattered electrons, and the final experimental results are
compared with calculations performed using distorted wave Born
approximation and the modified relativistic binary encounter Bethe
model.The sensitivity of the results on the atomic parameters is
explored. Observed agreements and discrepancies between the
experimental results and theoretical estimates, and their dependence
on the {\em specific} atomic parameters are reported.

\end{abstract}

\maketitle

%=====================================================================

%\begin{document}

%=====================================================================
\section{Introduction}
\label{sc:intro}

Importance of electron impact excitation and ionization data in
various materials analysis techniques such as electron probe
microanalysis (EPMA), Auger electron spectroscopy (AES), etc. need not
be overemphasized. Precise and accurate knowledge of the corresponding
cross sections is used as input either in the form of look-up
tables
or functional dependence on electron impact energies of the electron
probe used for such analysis. The inner shell ionization
probabilities, extracted from the above-mentioned data, are also
pivotal to many other material analysis techniques, apart from their
importance in understanding the physical process of ionization in
multi-electron bound systems~\cite{Powell1976}.

Inner shells of atoms can be excited by knocking off the bound
electrons to the continuum or unfilled quasi-bound orbitals. Vacancies
thus created are filled by the electrons from the outer  shells,
resulting in the emission of photons. In addition, migration of
vacancies
through Coster-Kronig(CK) transitions among different subshells
($L$-shell and above) as well as to other inner shells, leads to
photon emission with different energies and yields, which are
complicated by the fact that the corresponding transition
probabilities need to be accurately known. From the observation and
quantitative estimation of the related photon yield with high
precision, the inner shell ionization cross sections can be obtained,
in principle, utilizing the known or pre-determined parameters, such
as the fluorescence yield, CK transition probabilities and the
sublevel-specific radiative decay probabilities from experiments or
theoretical estimates. These important parameters are collectively
known as the atomic relaxation parameters.

The above mentioned relaxation parameters are obtained from
experiments or from theoretical estimates \cite{Bambynek1972} and are
available from various data bases. However, some of these parameters
are quoted with large uncertainties due to various processes involved.
For example, the fluorescence yield for a specific subshell depends on
the primary vacancy distributions, which in turn depends on the mode
of vacancy creation in the subshell. It is also expected that
migration of vacancies through CK transition would alter the primary
vacancy distributions and hence the fluorescence yield.

Photon emission by electron impact is also possible as a multistep
process through Auger transition, followed by creation of vacancy in
the inner subshells by virtual photons\cite{Chattarji1976}. The above
process involving virtual photons can only be accounted for by
invoking quantum electrodynamics and the associated electromagnetic
interaction between the bound electrons, which involve both
Coulomb interaction and the magnetic interaction due
to the moving electrons. In case of lighter elements, the motion of
inner shell electrons are in the non-relativistic regime ($v/c
\rightarrow 0$), and therefore, the quantum effects due to magnetic
interaction becomes negligible. Thus, inclusion of the Coulomb 
interaction alone in estimating the electron impact ionization cross 
sections results in reasonable agreement with the experimental results
for the lighter atoms. For heavier atoms like the ones considered in
this experiment, the magnetic interaction can no longer be ignored,
and related estimates of the electron impact ionization cross sections
should take magnetic interaction into account as well. Theoretical
estimates based on above has been done in recent times for Gold
(Au)\cite{Pindzola2015}.

Experiments on electron impact ionization which were done earlier,
were focused primarily on $ K $-shell ionization cross section, while
$L$ and $M$ shell ionization data were seldom reported\cite{Joy1995}.
One of the major problems faced in the interpretation of experimental
results based on established theories is that the extracted subshell
specific ionization cross sections do not agree with the theoretical
estimates for all the subshells. Recently, many authors have reported
$ L $ X-ray production cross sections for a few elements, and
validation of various theoretical models are done using the data.
Comparison between theory and experimental data on L-subshell
production cross sections in Gadolinium (Gd, $Z=64$) and Tungsten (W,
$Z=74$) were done by Wu et al.\cite{Wu2010_1}. Their experimental
results on $L_\alpha$ and $L_\beta$ lines agree reasonably well with
DWBA theory including exchange interaction for W but deviates by
15-20\% in case of Gd. Similar comparative studies were done by Varea
et. al \cite{Varea2011} on Hf, Ta, Re, Os, Au, Pb, and Bi, where
experimental results for $L_ {\alpha}$ and $L_{\beta}$ lines are
explained well by DWBA theory for Ta, Os, Au, Pb and Bi but are lower
than the theoretical estimates by $\sim 35\;\%$ for Hf and Re.

\begin{figure}
\begin{center}
\includegraphics[width=7cm,keepaspectratio=true] {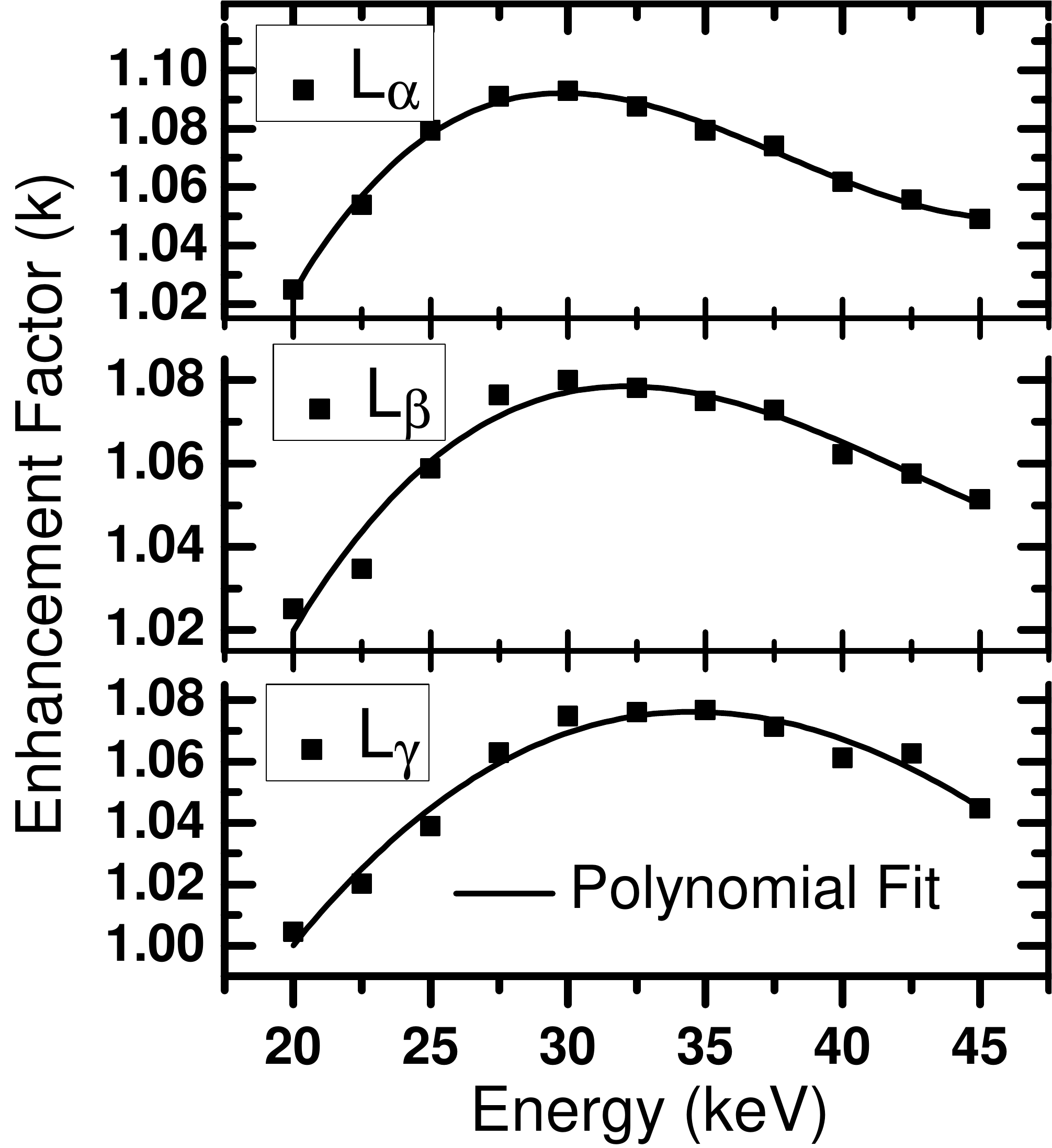}
 % backingcorr.jpg: 0x0 pixel, 0dpi, 0.00x0.00 cm, bb=
\end{center}
\caption{Enhancement factor due to the presence of Aluminum backing
 in Thorium target, as obtained from PENELOPE simulation}
\label{Fig:backingcorr}
\end{figure}

In the present work, the $L_{\alpha} , L_{\beta}, L_{\gamma}$
production cross sections in Lead and Thorium are measured, and the
results are converted to the subshell specific ionization cross
sections. Because of the finite thickness of the target materials,
single collision condition within the target has to be ensured. In
arriving at the ionization cross sections from the production cross
sections, corrections due to multiple collisions per beam traverse was
done using a Monte Carlo simulation procedure. Parameter dependence in
extracting the ionization cross sections are also explored to check
the sensitivity to parameter variations. The cross sections obtained
from experiment are compared with a)the theoretical results based on
the distorted wave Born approximation (DWBA) including relativistic
effects and exchange interactions into account \cite{bote2008}, as
obtained from the PENELOPE\cite{penelope2} code, and b) the modified
relativistic binary encounter Bethe (MRBEB)
~\cite{Guerra2012el,Guerra2015} model-based estimates. To the best of
our knowledge, the subshell specific ionization cross section for all
the $L$-subshells of Thorium are reported here for the first time at
the energy values near the corresponding ionization threshold.

%=====================================================================

\section{Experimental Details} 
\label{sc:expt}

The Experimental set-up consists of an in-vacuum energy dispersive
spectrometer with a focusable electron gun(up to 50 keV),
electrically cooled silicon PIN diode based X-ray detector, thin film
target holder, and Faraday cup. The X-ray detector was placed in the
meridian plane at $55^0$ with respect to the beam axis. The pressure
maintained inside the vacuum chamber was $~ 5\times10^{-7}$ mbar.
Details of the experimental arrangement are described elsewhere in
detail\cite {Rahangdale2014}.

The targets used in the experiment were made by using two different
techniques. Self-supporting Lead targets were made by electron beam
vapor deposition.The thickness of the thin film of Lead, deposited on
a glass substrate was monitored during deposition using a quartz
thickness monitor. Thorium targets were made by electro-deposition on
200~$\mu {\rm g/cm}^2$ thick Aluminum foil (99.99~\% purity). Electro-
deposition of Thorium oxide (ThO$_2$) on the foil cathode from a
Thorium nitrate solution in 2-propanol solvent, was monitored by
measuring the electrode current and the duration of deposition. Foil
thicknesses were measured by an alpha energy loss
spectrometer\cite{Rahangdale2014} and the measured thicknesses are
$78.1\,\pm\,3.8\,\,\mu{\rm g/cm}^2$~(Th) and
$82.0\,\pm\,4.2\,\,\mu{\rm g/cm}^2$~(Pb).

\begin{figure}
\begin{center}
\includegraphics[width=\columnwidth,keepaspectratio=true] {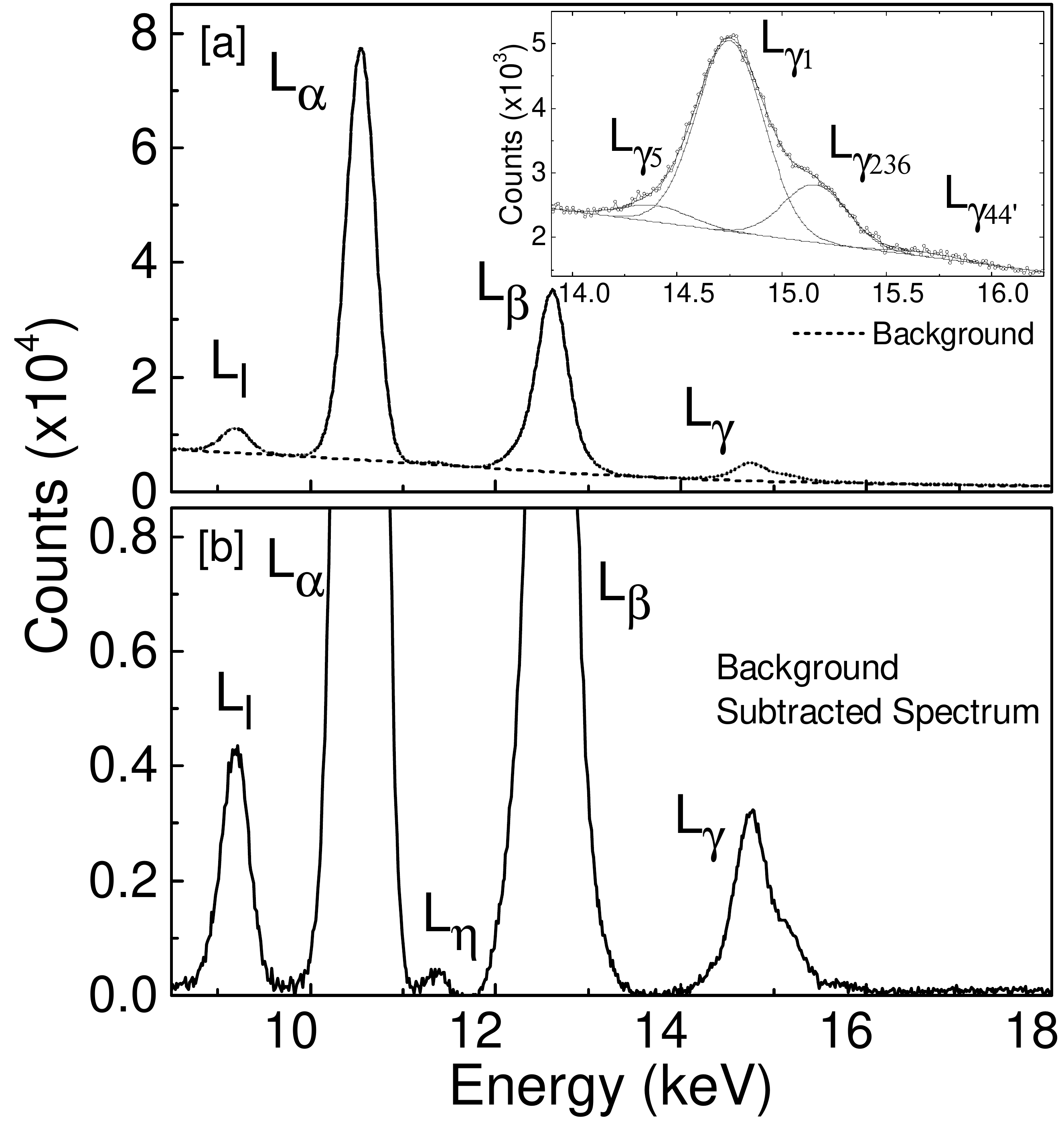}
\end{center}
\caption{Spectrum of Lead due to 35 keV electron impact. [a] Raw
Spectrum with fitted $L_\gamma$ are shown in the inset. [b] Same
spectrum after background subtraction.}
\label{Fig:Pb_Spectrum}
\end{figure}

X-rays generated due to electron impact were detected by X-ray
detector (model XR-100CR from Amptek, USA), having energy resolution
of 165 eV (FWHM) for 5.9 keV photons. Mylar foil of 100~$\mu$m
thickness was placed in front of the 25.4~$\mu$m Beryllium window to
reduce flux of $M$ X-rays.The efficiency of the detector was measured
by (i) $K$-shell ionization of Copper by electron impact and (ii)
using characteristic X-ray lines from a calibrated $^{241}$Am source.
The efficiency curve was fitted with equation $\epsilon(E)=1.58\times
10^{-5}+1.31\times 10^{-6}E-8.63\times 10^{-8}E^2$. The signal from
the detector was fed to a multi-channel analyzer (MCA) through a
shaping amplifier. For each data acquisition run, the count rates were
kept low ($<300$ counts/sec), so that there was no pile-up in the
detector and therefore, no dead time correction was required for the
MCA.

In the case of targets backed by the thick substrate, the electrons
can be back-scattered from the substrate material and re-enter the
target. These backscattered electrons can significantly change the
original X-ray yield. It is necessary to correct for this enhancement
of X-ray yield for the thick film backed targets like the Thorium
targets used in this work. Also for obtaining the accurate ionization
cross sections, one has to ensure that the target thickness is such
that the projectile electrons do not ionize the target atoms more than
once, thereby satisfying the single collision condition. A Monte Carlo
simulation based on PENELOPE\cite{penelope2} computer code was
performed to quantify the effect of electron back-scattering on the
measured X-ray yield and thereby, ensure the inclusion of single
collision events.

The Monte Carlo simulation code PENELOPE is a versatile program for
estimation of the effects of electron-photon transport in materials.
The main advantage of invoking a Monte Carlo simulation at this stage
is: 1) ease of incorporating sophisticated interaction models and
2)
convenience and capability of including arbitrary geometry into the
calculation. However, in achieving this level of sophistication, the
crystalline structure of the solid materials are completely ignored by
considering the interacting media as homogeneous, isotropic and
amorphous with definite composition and density. It is evident that
such a {\em gas-like} model of the medium may be considered as a valid
approximation for electron beam interaction with thin films of solids,
however, the simulation results are likely to deviate from reality for
thicker solid interaction media.

PENELOPE uses the combination of numerical and analytical physical
interaction models to track down the encounter of electrons and
photons with matter. Specifically, the effect of electron impact inner
shell ionization is taken into account from the numerical differential
cross sections (DCS)\cite{Bote2009871} obtained from DWBA based
calculations\cite{bote2008}.

%\bibitem{Bote2}{D. Bote and F. Salavat, Phys Rev {\bf A 77}, 04270 (2008)}
\begin{figure}
\begin{center}
\includegraphics[width=\columnwidth,keepaspectratio=true] {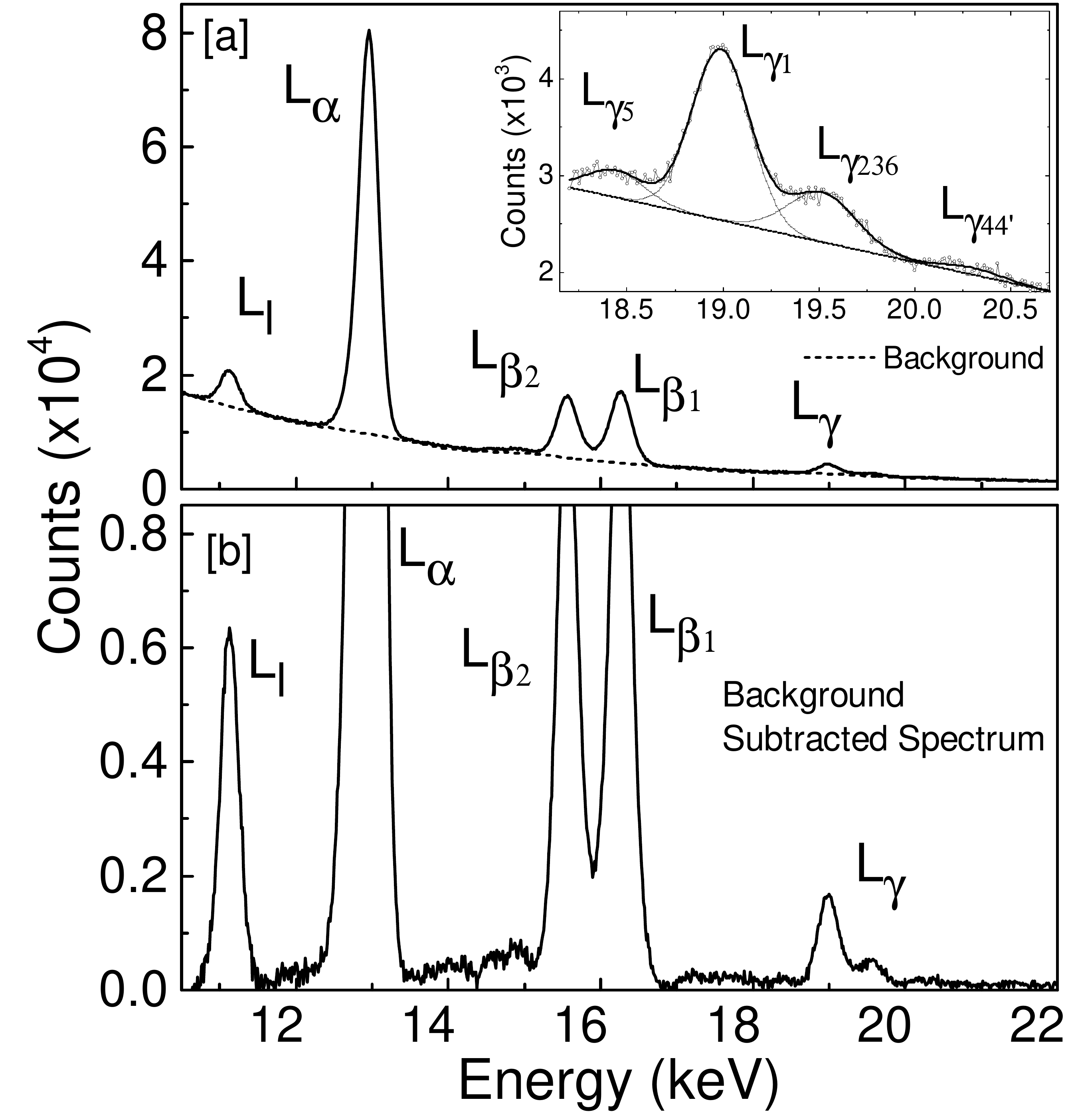}
\end{center}
\caption{Spectrum of Thorium due to 35 keV electron impact. [a] Raw
Spectrum with fitted $L_\gamma$ are shown in the inset. [b] Same
spectrum after background subtraction.}
\label{Fig:Th_Spectrum}
\end{figure}
While generating simulated X-ray spectra by PENELOPE, the process
becomes inefficient and time consuming due to 1) low inner shell
ionization and subsequent radiative decay probabilities, and 2) use of
thin film media in the experiment. This results in large variance and
reduces the predictive power of simulation. To reduce the time spent
on computation and to increase the efficiency, it is necessary to use
a variance reduction technique, known as interaction forcing. PENELOPE
implements the process by artificially reducing the mean free path
relevant to the process, but keeping the probability distribution
functions for energy loss and angular deflections the same as for the
real process. Finally, the biasing introduced by the simulation
process is corrected for by applying appropriate statistical
weights\cite{penelope2}.

% \bibitem{penelope2}{F. Salavat, J. M Fern{\'a}ndez-Varea and J. Sempau, PENELOPE 2011: a code system for 
% Monte Carlo electron and photon transport, NEA Data Bank, NEA/NSC/DOC(2011)5.}      

Simulation was carried out with a pencil-like electron beam of 2~mm diameter 
impinging on a thin target at normal incidence. Entry of the projectile electrons 
into the target, resulting ionization events and emission of X-rays were 
recorded event by event. From the simulation of a large sample of events, 
the maximum probability of inner shell ionization per projectile electron was 
$\sim 0.53$ for both the target films used in the experiment. This
number,
being less than unity, ensures that single collision condition was satisfied 
in the experiment.

To account for the effect of electron back-scattering in the Thorium target, 
simulation was done with and without aluminum backing. It was found that up to 
$\sim 4$\% of the electrons which ionized the Thorium atoms and subsequently 
generated $L$ X-rays were back-scattered from aluminum. The back-scatter 
fraction, however, was found to depend on electron 
energy ($E$). After obtaining X-ray yield with good statistics from  
simulation, the corresponding enhancement factor $k(E)$ was obtained as:
\begin{equation}
k(E)=\frac{\rm Counts\,\,under\,\,{\it L_x}\,\,peak\,\,for\,\,Al\,\,backed\,\,target}
          {\rm Counts\,\,under\,\,{\it L_x}\,\,peak\,\,for\,\,unbacked\,\,target},
\end{equation}
where $x$ is $\alpha$, $\beta$ or $\gamma$.

The $k(E)$-values, obtained from simulation, are plotted in the
Fig.~\ref{Fig:backingcorr} for the $L$ X-rays. The values of $k(E)$
lies in the range: 1.005 to 1.095.

%=====================================================================

\section{Data Analysis}
\label{sc:data}

The X-ray spectra of Lead and Thorium, resulting from electron
bombardment at 35 keV, are shown in the
Figs.~\ref{Fig:Pb_Spectrum}~and~\ref{Fig:Th_Spectrum} respectively.
Individual $L_x$ peaks in the observed spectra were fitted with
Gaussian profile over the bremsstrahlung background, as shown in the
figures, to obtain the corresponding X-ray yield ($N_X$). The
bremsstrahlung background over the region of interest was considered
as linear due to small interval of energy spanned by each peak. $L_l$
and $L_\alpha$ peaks were fitted with single Gaussian functions for
both Lead and Thorium. $L_ \beta$ peaks in Thorium could be resolved
in $L_ {\beta1}$ and $L_{\beta2}$, and were fitted with two Gaussian
profiles. The $L_\beta$ peaks could not be resolved for Lead and
therefore, a single Gaussian with higher FWHM value was fitted. $L_
\gamma$ peaks in both the targets were resolved into three constituent
lines viz., $L_{\gamma_5}$, $L_{\gamma_1}$, and $L_{\gamma_{236}}$.
Fitted $L_ \gamma$ spectra are shown in the insets of Figs
~\ref{Fig:Pb_Spectrum}(a) and ~\ref{Fig:Th_Spectrum}(a). The net
counts obtained  by fitting the spectra were corrected for self-
absorption due to finite target thicknesses, assuming the oblique path
of the X-rays through the target materials.

The X-ray production cross sections were obtained from the
measurements using the formula:
\begin{equation}
\label{cross}
\sigma_i(E)=\frac{N_X A}{\epsilon (E') t  N_e N_A k(E)},
\end{equation}
where, $\sigma_i(E)$ is the production cross section of $L_x$ line at
projectile electron energy $E$, $N_X$ is the net yield of X-rays after
self-absorption correction during a time interval $T$, $\epsilon
(E')$ is the effective efficiency of detector at  photon energy $E'$
which is the energy of $L_x$ line centroid,
$t$ is the thickness of target, $A$ is the  mass number of the target
material, $N_e$ is the total number of electrons impinging on the
target during the same interval $T$, $N_A$ is the Avogadro number and
$k(E)$ is the enhancement factor defined as above($k=1$ for Lead). The
effective efficiency of the detector includes the effect of the
geometric factor, attenuation due to Mylar and the intrinsic
efficiency of the detector.

The experimentally obtained production cross sections were converted
to the ionization cross sections using
Eqs.~\ref{L1},~\ref{L2},~\ref{L3}, given as:

\begin{equation}
\label{L1}
\sigma_{L_1}=\frac{\sigma_{L_{\gamma_{2+3}}}}{\omega_1
S_{\gamma_{2+3},1}},
\end{equation}

\begin{equation}
\label{L2}
\sigma_{L_2}=\frac{\sigma_{L_{\gamma_{1+5}}}}{\omega_2S_{\gamma_{1+5},2
}} -\sigma_{L_1}f_{12},
\end{equation}

\begin{equation}
\label{L3}
\sigma_{L_3}=\frac{\sigma_{L_{\alpha}}}{\omega_3 S_{\alpha
,3}}-\sigma_{L_1}(f_{12}f_{23}+f_{13})-\sigma_{L_2}f_{23}.
\end{equation}

$S_{i,I}$ is the fraction of radiative transition resulting from
vacancy created in the $I^{th}$ subshell associated with the $L_i$
peak, $\omega_i$ is the fluorescence yields corresponding to subshells
$L_i$, and $f_{ij}$ is the Coster-Kronig transition probability
between the $L_i$  and $L_j$ subshells. The production cross sections
corresponding to the $L_{\gamma_{1}}$, $L_{\gamma_{236}}$ and
$L_\alpha$ transitions are used in the above equations, which is the
recommended combination (see ref.~\cite{Lapicki2005}), among  many
other combinations, to find the ionization cross sections. The 
atomic
relaxation parameters, used in the calculations, are taken from the
Refs.~\cite{Campbell2003}~and ~\cite{Campbell1989}.
Tables~\ref{tab:fluor}~and~\ref{tab:rady} enlist all the parameters
used in this work.

\begin{table}[h]
\begin{center}
\begin{tabular}{lcccccc}
\hline 
Target  & $\omega_1$ & $\omega_2$ & $\omega_3$ & $f_{12}$ & $f_{13}$ &
$f_{23}$
\\ 
\hline 
Lead & 0.1	 & 0.397	 & 0.343	 & 0.064	& 
0.61 & 0.119\\
Thorium & 0.17 & 	0.503 & 	0.424 & 	0.06	& 
0.66	& 0.103\\
\hline 
\end{tabular}
\end{center}
\caption{Fluorescence yield and Coster-Kronig transition 
probabilities used in this work.}
\label{tab:fluor}
\end{table}

% \onecolumngrid
\begin{table*}
\begin{tabular}{lllllllllll}
\hline \hline
Line    & \begin{tabular}[c]{@{}l@{}}Source \\ shell\end{tabular} &
\begin{tabular}[c]{@{}l@{}}Vacant\\  shell\end{tabular} &  &
\begin{tabular}[c]{@{}l@{}}Transition \\ Energy (keV)\end{tabular} &
\begin{tabular}[c]{@{}l@{}}Radiative \\ Yield ($\Gamma$)\end{tabular}
& \begin{tabular}[c]{@{}l@{}}Radiative Yield \\ Fraction
($S_{i,I}$)\end{tabular} &  & \begin{tabular}[c]{@{}l@{}}Transition\\
Energy (keV)\end{tabular} & \begin{tabular}[c]{@{}l@{}}Radiative\\
Yield ($\Gamma$)\end{tabular} & \begin{tabular}[c]{@{}l@{}}Radiative
Yield \\ Fraction ($S_{i,I}$)\end{tabular} \\
\hline
 & & & & \multicolumn{3}{l}{-----------------Lead------------------ }
 & & \multicolumn{3}{l}{---------------Thorium---------------} \\
\hline
$l_l$            & M1 & L3 & & 9.184  & 0.085 & 0.0406 &  & 11.119  & 0.146 & 0.0449  \\
$l_{\alpha_2}$   & M4 & L3 & & 10.449 & 0.164 & 0.0786 &  & 12.81   & 0.250 & 0.0765  \\
$l_{\eta}$       & M1 & L2 & & 11.347 & 0.052 & 0.0216 &  & 14.507  & 0.084 & 0.0215  \\
$l_{\beta_6}$    & N1 & L3 & & 12.141 & 0.021 & 0.0101 &  & 14.973  & 0.037 & 0.0115  \\
$l_{\beta_2}$    & N5 & L3 & & 12.623 & 0.293 & 0.1402 &  & 15.621  & 0.474 & 0.1451  \\
$l_{\beta_4}$    & M2 & L1 & & 12.304 & 0.456 & 0.3458 &  & 15.64   & 0.756 & 0.3588  \\
$l_{\beta_1}$    & M4 & L2 & & 12.614 & 1.884 & 0.7808 &  & 16.202  & 2.951 & 0.7598  \\
$l_{\beta_{15}}$ & N4 & L3 & & 12.601 & 0.032 & 0.0155 &  & 15.588  & 0.051 & 0.0158  \\
$l_{\beta_5}$    & O4 & L3 & & 13.013 & 0.042 & 0.0204 &  & 16.211  & 0.099 & 0.0305  \\
$l_{\beta_5}$    & O5 & L3 & & 13.013 & 0.042 & 0.0204 &  & 16.211  & 0.099 & 0.0305  \\
$l_{\beta_3}$    & M3 & L1 & & 12.791 & 0.501 & 0.3796 &  & 16.423  & 0.696 & 0.3303  \\
$l_{\gamma_5}$   & N1 & L2 & & 14.305 & 0.013 & 0.0056 &  & 18.361  & 0.022 & 0.0058  \\
$l_{\gamma_1}$   & N4 & L2 & & 14.762 & 0.404 & 0.1677 &  & 18.982  & 0.685 & 0.1765  \\
$l_{\gamma_2}$   & N2 & L1 & & 15.099 & 0.120 & 0.0909 &  & 19.302  & 0.207 & 0.0983  \\
$l_{\gamma_3}$   & N3 & L1 & & 15.215 & 0.145 & 0.1103 &  & 19.503  & 0.218 & 0.1036  \\
$l_{\gamma_6}$   & O4 & L2 & & 15.176 & 0.054 & 0.0227 &  & 19.596  & 0.133 & 0.0343  \\
$l_{\gamma_4}$   & O3 & L1 & & 15.775 & 0.052 & 0.0396 &  & 20.289  &
0.101 & 0.0478  \\
$l_{\gamma_4'}$  & O2 & L1 & & 15.755 & 0.052 & 0.0396 &  & 20.289  &
0.101 & 0.0478  \\
\hline
\end{tabular}
\caption{Radiative yields for Lead and Thorium, from Campbell and 
Wang\cite{Campbell1989}.}
\label{tab:rady}
\end{table*}

% \twocolumngrid
It is evident from Eq.~\ref{L2} that the $L_{\gamma_{2+3}}$ production
cross section is needed to obtain the $L_1$ sub-shell ionization cross
section. However, it is not directly available from experiment due to
the limited resolution of the X-ray detector. As mentioned earlier,
the $L_\gamma$ peak is resolved into  $L_{\gamma_5}$, $L_{\gamma_1}$,
and $L_{\gamma_{236}}$ lines. Therefore, the  production  cross
section of $L_{\gamma_{2+3}}$ line is obtained by subtracting  the
contribution of $L_{\gamma_6}$ from the experimentally obtained
$L_{\gamma_{236}}$ peak. The contribution of $L_{\gamma_6}$, in turn,
is obtained from the ratio: $\Gamma_{\gamma_6}/{\Gamma_{\gamma_1}}$
and
the $L_{\gamma_1}$ peak counts of the fitted spectrum.

The $L_\gamma$ peaks were not observed at 16 keV electron impact
energy for Lead, and at 20 and 22.5 keV energy for Thorium. Also at 25
keV electron beam energy, only $L_{\gamma_1}$ could be observed for
Thorium and therefore, only the $L_2$ and $L_3$ ionization cross
sections could be obtained. $L_1$, $L_2$ and $L_3$ ionization cross
sections were extracted from the data at all energies above 16 keV for
Lead and 25 keV for Thorium.

The $L_\beta$ line of Thorium was resolved into $L_{\beta_1}$ and
$L_{\beta_2}$ peaks in the obtained spectra. To cross-check and verify
the obtained ionization cross sections, attempts were made to extract
the $L_1$ and $L_2$ ionization cross sections from the $L_{\beta_1}$
and $L_{\beta_2}$ production cross sections using the equations
~\ref{Lb1} and ~\ref{Lb2}~\cite{Shima1981}.

\begin{widetext}
\begin{eqnarray}
\label{Lb1} 
\sigma_{(L_{\beta_1}+L_{\beta_5}+L_{\beta_3})}&=&S_{\beta_
5,3}\omega_3\sigma_{L_3} + \left[S_{\beta_
1,2}\omega_2+S_{\beta_5,3}\omega_3f_{23}
\right]\sigma_{L_2} \nonumber \\ &&+ \left[S_{\beta_1,
2}f_{12}\omega_2+S_{\beta_5,3}\omega_3(f_{13}+f_{12}f_{2
3}) +S_{\beta_3,1}\omega_1\right]\sigma_{L_1},
\end{eqnarray}
\begin{eqnarray}
	\label{Lb2}
\sigma_{(L_{\beta_2}+L_{\beta_6}+L_{\beta_4})}&=&S_{\beta_{2+6},3}
\omega_3\sigma_{L_3} + S_{\beta_{2+6},3}\omega_3 f_
{23}\sigma_{L_2}\nonumber \\ &&+\left[S_{\beta_{2+6},3}\omega_3(f_
{13}+f_ {12}f_{23})+S_{\beta_4,1}\omega_1\right]\sigma_ {L_1}
\end{eqnarray}
\end{widetext}

where, the symbols used have the usual meaning, as explained for
equations~(~\ref{L1},~\ref{L2} and~\ref{L3}). The same set of atomic
relaxation parameters was used. Ionization cross sections for $L_3$
subshell, needed as input, were obtained from the DWBA estimates. A
good reason for using the theoretical estimates for $L_3$ subshell is
that the experimental results are found to be in reasonable agreement
(see Fig.~\ref{Fig:Th_Ionization}).

The $L_2$ subshell ionization cross sections, obtained as above for
the given energy range, were found to be in good agreement with the
results obtained from the $L_\alpha$ and $L_\gamma$ cross sections
(see Eqs.~\ref{L1}, \ref{L2} and \ref{L3}). However, the calculated
$\sigma_{L_1}$ values were not at all consistent with the
corresponding results. Minor changes in the relaxation parameters
within the allowed range of variation (see Ref.~\cite{Campbell2003})
restores $\sigma_{L_1}$ values to come closer to the previously
obtained results (see Fig.~\ref{Fig:Th_Ionization}), without causing
much deviation in $\sigma_{L_2}$ values. This indicates the need for
possible modification of the atomic relaxation parameters.

%=====================================================================
%=====================================================================

\section{Results and Discussion}
\label{sc:res}

The X-ray production cross sections, determined from the experiment
and the ionization cross sections, obtained from the experimental
data, are shown in Tables \ref{Tab:Pb_crosssections} and
\ref{Tab:Th_crosssections} for Lead and Thorium respectively. The
uncertainties in the cross section values are indicated. Overall
uncertainties for $L$ X-rays production cross sections are $\sim 11 -
12\%$ for both the elements. Contribution to  the uncertainties are
from 1) detector efficiency ($\sim 10$\%), 2) target thickness
measurement ($\sim 5$\%) and 3) beam current measurement ($\sim$ 3\%).
Considering propagation of errors as per Eqs. (\ref{L1}, \ref{L2},
\ref{L3}), the uncertainties in the corresponding ionization cross
sections are $\sim 20\%$, and including the uncertainties in the
relaxation parameters within their quoted ranges, the errors in the
ionization cross sections are larger $ \sim 30\%$.

\begin{table*}
\centering
\begin{tabular}{cccccccc}
\hline \hline
Energy &
\multicolumn{3}{c}{Production cross section} &       &
\multicolumn{3}{c}{Ionization cross section}
\\ 
\hline
& $ L_\alpha $ & $ L_\beta $ & $ L_\gamma $ & & $ L_{1} $ & $  L_{2} $
& $L_{3}$ \\ (KeV) & (barn) & (barn) & (barn) &       & (barn) &
(barn) & (barn) \\
\hline
16 &  43.5(5.2) &  9.9(1.2)  & ..       &\hspace{4mm}  & .. & .. & 164.5(20.6) \\
18 &  55.3(6.7) & 21.7(2.6)  & 1.4(0.2) &\hspace{4mm}  &  3.9(2.2) &  21.9(4.0) & 204.4(26.3) \\
20 & 117.7(14.2) & 52.6(6.4)  & 4.5(0.5) &\hspace{4mm}  & 22.8(8.7) &  64.6(11.6) & 423.7(56.1) \\
23 & 129.1(15.6) & 63.2(7.6)  & 6.1(0.7) &\hspace{4mm}  & 44.6(14.7) &  84.0(15.4) & 451.0(62.0) \\
25 & 137.3(16.6) & 70.1(8.5)  & 6.8(0.8) &\hspace{4mm}  & 49.7(16.4) &  93.7(17.2) & 477.6(66.0) \\
28 & 146.6(17.7) & 76.8(9.3)  & 8.3(1.0) &\hspace{4mm}  & 63.4(20.6) & 113.4(20.7) & 502.0(70.8) \\
30 & 159.0(19.2) & 83.8(10.1)  & 8.9(1.1) &\hspace{4mm}  & 72.3(23.1) & 119.9(22.1) & 542.6(76.8) \\
33 & 154.3(18.6) & 84.8(10.2)  & 9.0(1.1) &\hspace{4mm}  & 76.1(23.9) & 119.8(22.1) & 522.6(74.7) \\
35 & 154.0(18.1) & 90.2(10.9)  & 9.0(1.1) &\hspace{4mm}  & 70.4(22.6) & 121.7(22.2) & 524.6(74.4) \\
38 & 151.6(18.3) & 84.0(10.1)  & 8.7(1.1) &\hspace{4mm}  & 74.6(23.4) & 116.4(21.6) & 513.8(73.4) \\
40 & 147.5(17.8) & 81.6(9.8)  & 9.2(1.1) &\hspace{4mm}  & 80.9(25.0) & 121.9(22.1) & 493.8(71.8) \\
\hline \hline
\end{tabular}
\caption{Experimental Production and Ionization cross sections of Pb.}
\label{Tab:Pb_crosssections}
\end{table*}

\begin{table*}
\centering 
\begin{tabular}{cccccccc}
\hline \hline
 Energy & \multicolumn{3}{c}{Production cross section} &       &
 \multicolumn{3}{c}{Ionization cross section}
\\ 
\hline
& $ L_\alpha $ & $ L_\beta $ & $ L_\gamma $ & & $ L_{1} $ & $  L_{2} $ & $L_{3}$ \\ 
(KeV) & (barn) & (barn) & (barn) &       & (barn) & (barn) & (barn) \\
\hline 
20   &  20.5(2.0) & 02.8(0.3) &    ..    &\hspace{4mm} & ..         & ..         & 064.6( 8.0) \\
22.5 &  46.8(5.6) & 10.3(1.0) &    ..    &\hspace{4mm} & ..         & ..         & 147.5(18.4) \\
25   &  68.2(8.1) & 21.9(2.0) & 2.1(0.3) &\hspace{4mm} & ..         & 19.8(3.2) & 213.0(26.8) \\
27.5 &  87.6(10.5) & 31.5(3.1) & 5.6(0.7) &\hspace{4mm} & 29.2(7.3) & 41.6(7.1) & 252.2(34.7) \\
30   & 103.8(12.5) & 44.6(3.9) & 6.2(0.8) &\hspace{4mm} & 32.0(8.0) & 46.5(7.9) & 301.1(41.1) \\
32.5 & 113.4(13.6) & 51.7(4.5) & 6.9(0.8) &\hspace{4mm} & 33.9(8.4) & 52.3(8.9) & 329.2(44.8) \\
35   & 114.9(13.8) & 53.9(4.7) & 7.7(0.9) &\hspace{4mm} & 36.8(9.3) & 58.0(9.8) & 331.7(45.5) \\
37.5 & 122.1(14.6) & 56.7(5.0) & 8.0(1.0) &\hspace{4mm} & 36.5(9.1) & 61.6(10.4) & 354.0(48.3) \\
40   & 129.5(15.5) & 62.6(5.4) & 8.9(1.1) &\hspace{4mm} & 39.8(10.1) & 68.9(11.6) & 374.5(51.3) \\
42.5 & 124.9(15.0) & 59.8(5.3) & 9.5(1.2) &\hspace{4mm} & 41.0(10.3) & 74.0(12.4) & 358.5(49.5) \\
45   & 127.9(15.3) & 62.9(5.5) & 8.9(1.1) &\hspace{4mm} & 42.1(10.4) & 68.0(11.6) & 368.0(50.6) \\
\hline \hline 
\end{tabular} 
\caption{Experimental Production and Ionization cross sections of
Th.}
\label{Tab:Th_crosssections} \end{table*}

The experimental results are compared with the two different
theoretical estimates based on two different formalisms: 1) MRBEB
theory and 2) DWBA formalism. The DWBA theory based analytical
formulas for calculating the ionization cross sections for electron or
positron impact is given by Bote et. al. \cite{Bote2009871}. The
details of the MRBEB theory involved in these estimates can be found
in Ref.~\cite{Rahangdale2014,Guerra2012el,Guerra2015}.
The theoretical $L$-shell ionization cross sections, obtained from
these formalisms, are converted into production cross sections using
Eqs.~\ref{L3},~\ref{Lb}, and ~\ref{Lg} along with the relevant
relaxation parameters ~\cite{Shima1981}.

\begin{eqnarray}
\label{Lb}
\sigma_{L_{\beta}}&=&\sigma_{L_1}[\omega_1S_{\beta,1}+\omega_2f_{12}S_
{\beta,2} +\omega_3(f_{13}+f_{12}f_{23})S_{\beta,3}]\nonumber \\
&&+\sigma_{L_2}(\omega_2S_{\beta,2}+\omega_3f_{23}S_{\beta,3})+\sigma_
{L_3}\omega_3S_{\beta,3}
\end{eqnarray}

\begin{eqnarray}
\label{Lg}
\sigma_{L_{\gamma}}&=&\sigma_{L_1}[\omega_1S_{\gamma,1}+\omega_2f_
{12}S_ {\gamma,2}] +\sigma_{L_2}\omega_2S_
{\gamma,2}
\end{eqnarray}

\begin{figure}
\centering
{
\includegraphics[width=\columnwidth,keepaspectratio=true] {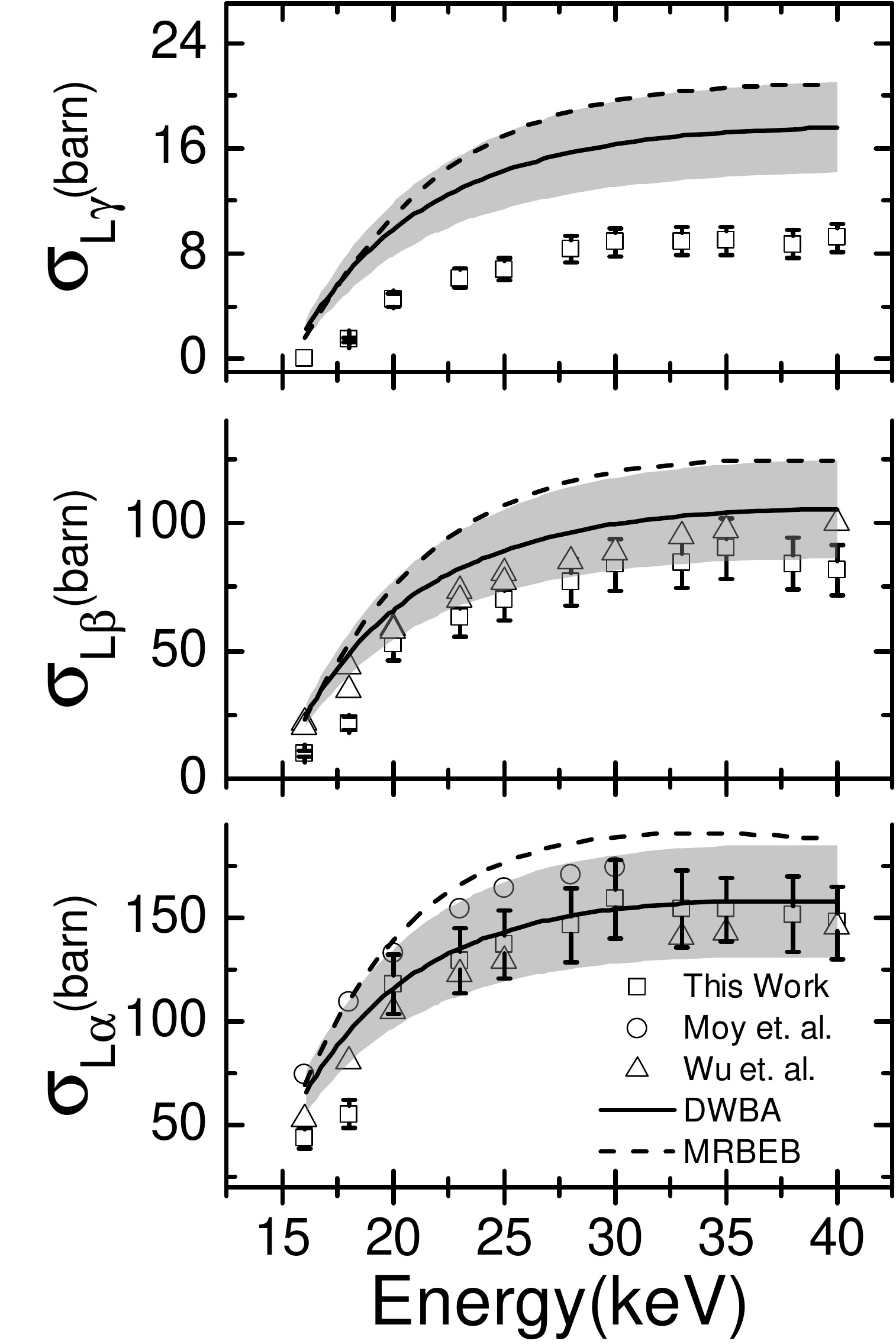}
\caption{Production cross section of $L_\alpha$, $L_\beta$ and
$L_\gamma$ lines of Pb. Theoretical curves are obtained using
relaxation parameters from \cite{Campbell2003} and
\cite{Campbell1989}. The shaded area is due to the uncertainty in the
adopted relaxation parameters.}
\label{Fig:Pb_Production}
}
\end{figure}

While comparing with theory, it should be noted that the relaxation
parameters, which are used to obtain theoretical production cross
sections, can themselves have uncertainties $\lesssim 50\%$. Table
\ref{tab:relpar} shows the recommended uncertainties in
Ref.\cite{Campbell2003}, which are adopted in this work.

\begin{table}
\begin{center}
\begin{tabular}{lccccccc}
\hline \hline 
Parameters & $\omega_1$ & $\omega_2$ & $\omega_3$ & $f_{12}$ &
$f_{13}$ & $f_{23}$ & $\Gamma$,s \\
$\%$ Error & 20 & 5 & 5	 & 50	& 15 & 10 & 10 \\
\hline 
\end{tabular}
\end{center}
\caption{Adopted errors in relaxation parameters.}
\label{tab:relpar}
\end{table}

\begin{figure}
\centering
{
\includegraphics[width=\columnwidth,keepaspectratio=true] {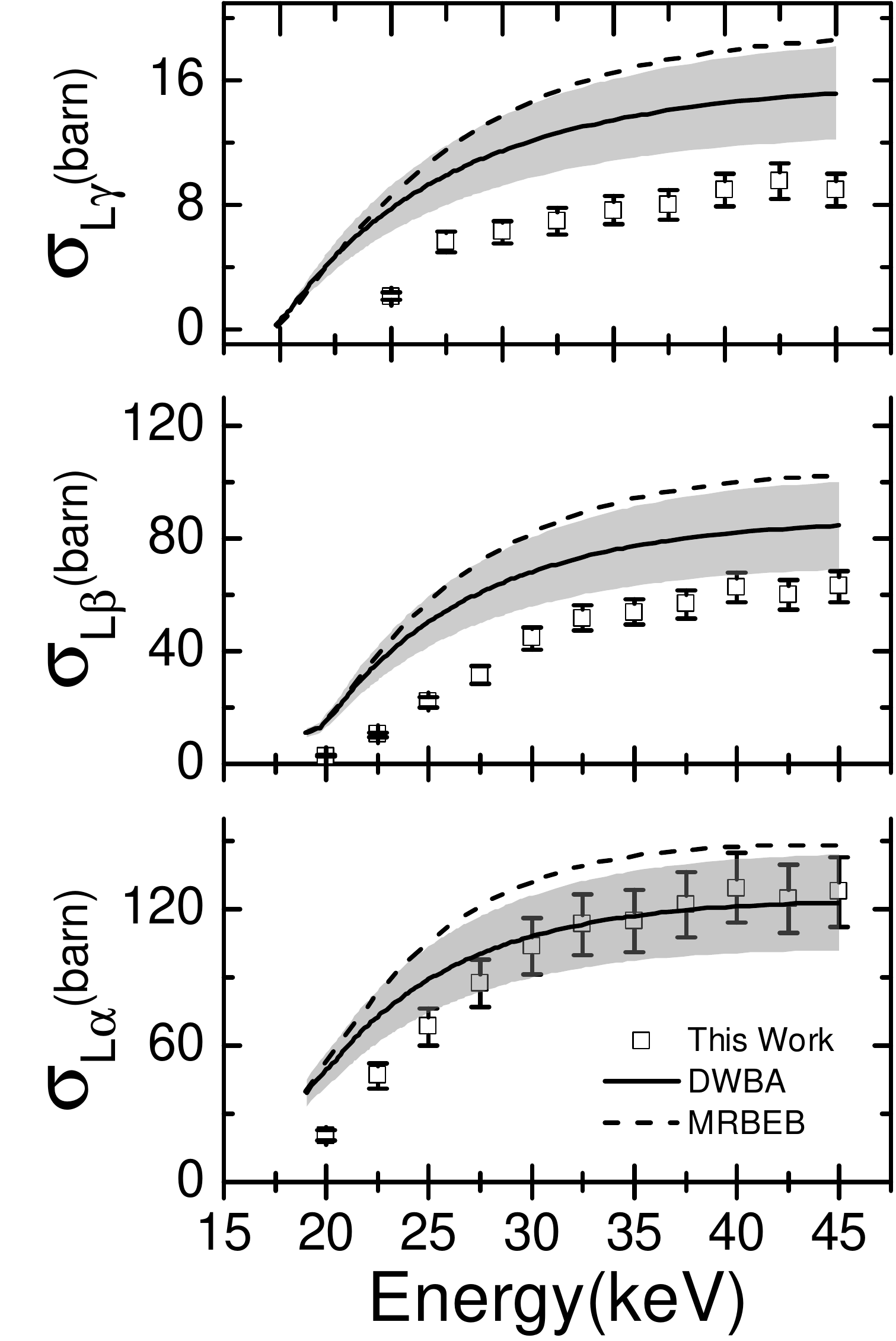}
\caption{Production cross section of $L_\alpha$, $L_\beta$ and
$L_\gamma$ lines of Th. Theoretical curves are obtained using
relaxation parameters from
\cite{Campbell2003} and \cite{Campbell1989}. The shaded area is due to
the uncertainty in the adopted relaxation parameters.}
\label{Fig:Th_Production}
}
\end{figure}

The $ L $ X-ray production cross sections of Lead and Thorium are
plotted in the Figures~\ref{Fig:Pb_Production} and
~\ref{Fig:Th_Production} respectively. Corresponding theoretical
estimates, based on the DWBA and the MRBEB theories are also plotted
on the same graphs. The shaded regions around the DWBA estimates in
both the graphs indicate the predicted uncertainty bands arising from
the uncertainties in the adopted relaxation parameters.

In case of Lead, the $L_\alpha$ and $L_\beta$ X-ray production cross
sections, based on measurements done by Wu et al.\cite{Wu2007} and Moy
et al.\cite{Moy2013}, are also shown in the
Fig.~\ref{Fig:Pb_Production}. These two sets of measurements are in
good agreement with our corresponding results. The DWBA estimates for
the $L_\alpha$ production cross sections of both the elements are in
good agreement with all three experimental data sets. The DWBA
estimates for $L_\beta$ lines of Lead overpredict the production cross
sections across the energy range of interest, but the estimates agree
with the experimental results within the predicted uncertainty band.
Considering the systematic trend in the experimental data over the
energy range, the results of Wu et al.\cite{Wu2007} are in better
agreement within the uncertainty band. Our results for Lead are
systematically on the lower end of the predicted band. The MRBEB
theory predicts larger production cross sections in all the cases,
with values grazing the upper end of the predicted uncertainty band of
DWBA estimates.

No other measurement of the $L$ X-ray production cross sections of
Thorium exists to the best of our knowledge. Our results agree with
the DWBA estimates for the $L_\alpha$ line (see
Fig.~\ref{Fig:Th_Production}). Comparison with DWBA estimates for the
$L_\beta$ and $L_\gamma$ production cross sections of Thorium indicate
the similar trend as that in Lead.

\begin{figure}
\centering
\includegraphics[width=\columnwidth,keepaspectratio=true] {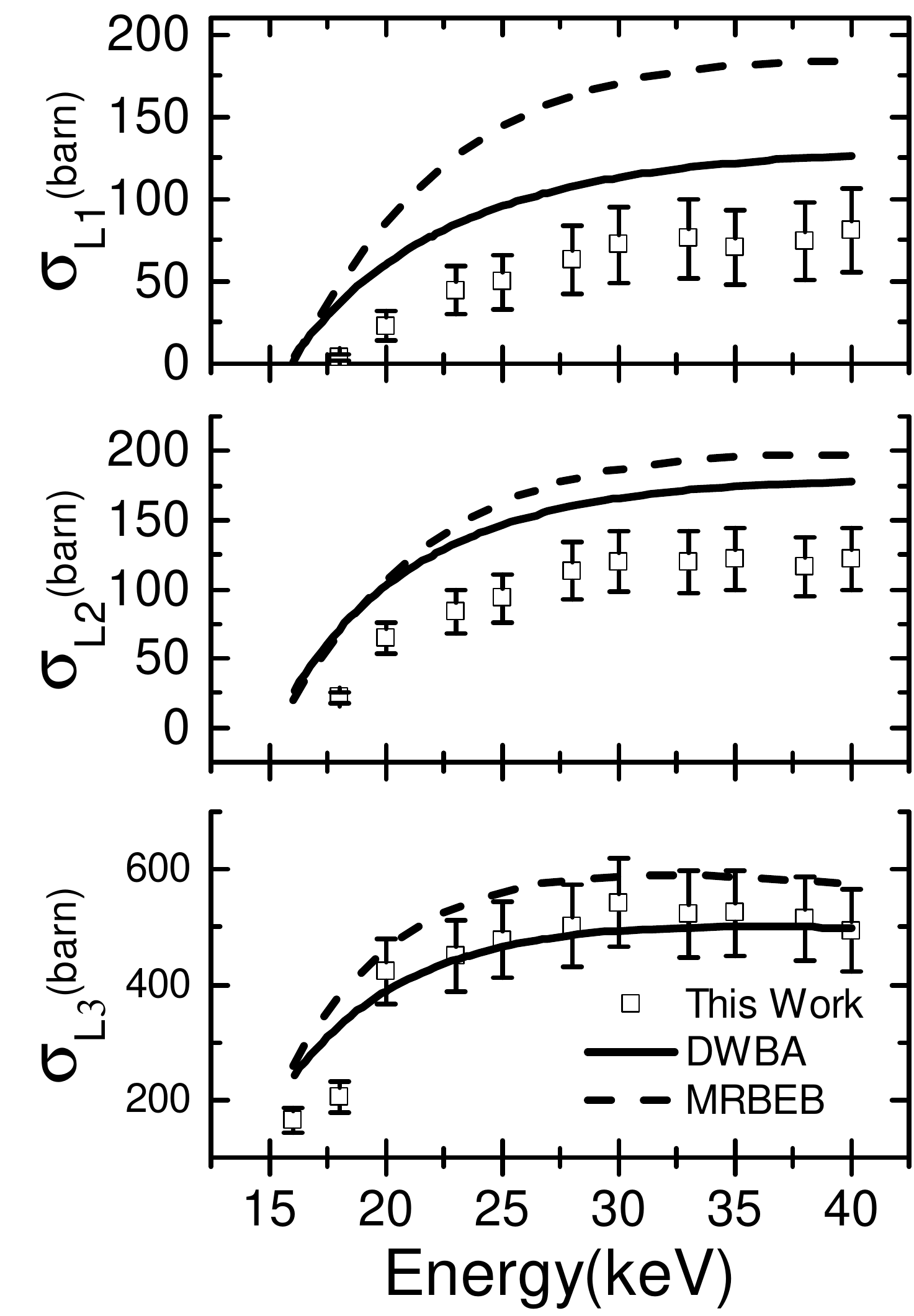}
\caption{$L_1$, $L_2$ and $L_3$ subshell ionization cross sections 
of Lead.}
\label{Fig:Pb_Ionization}
\end{figure}

\begin{figure}
\centering
{
\includegraphics[width=\columnwidth,keepaspectratio=true] {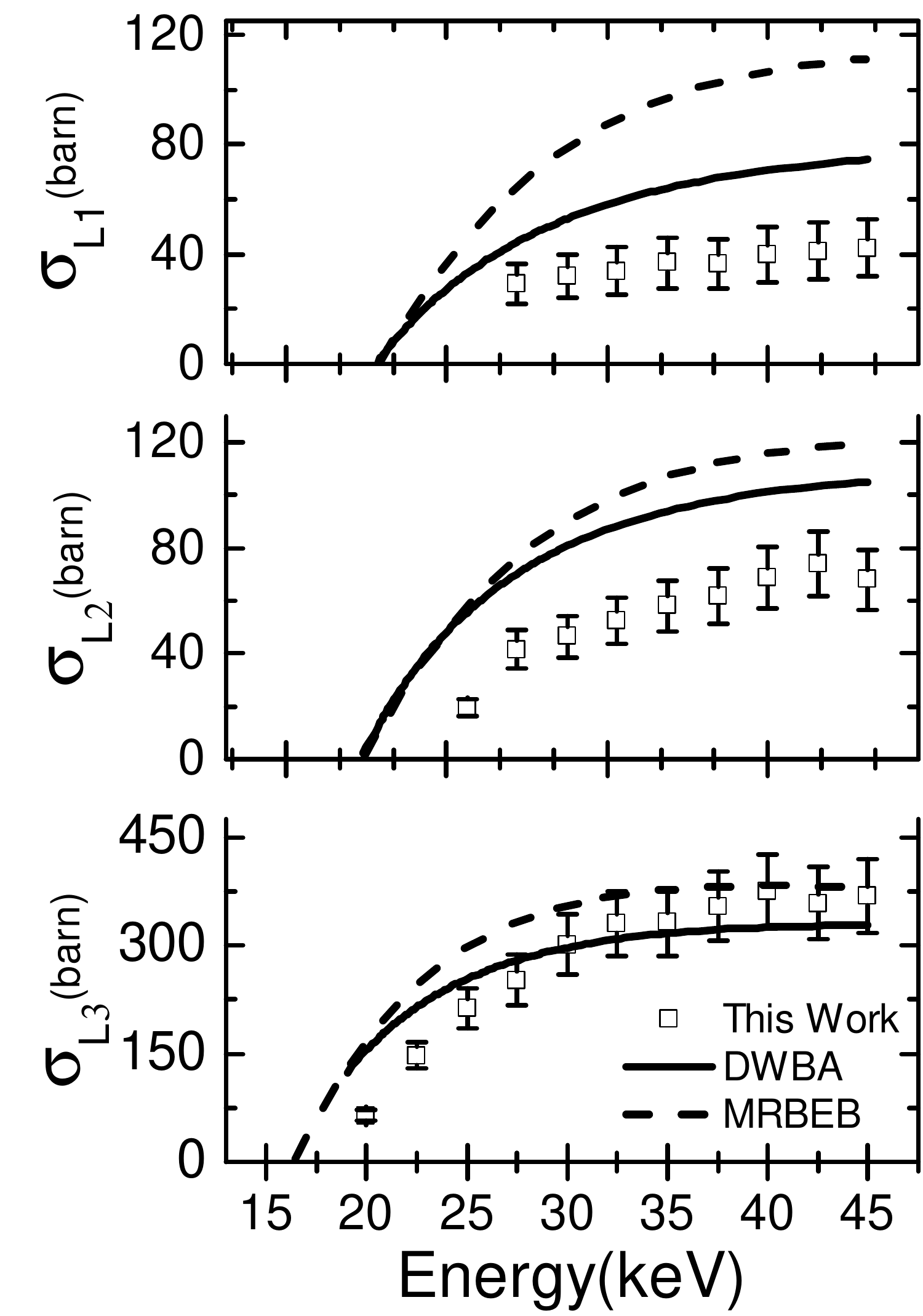}
\caption{$L_1$, $L_2$ and $L_3$ subshell ionization cross sections 
of Thorium.}
\label{Fig:Th_Ionization}
}
\end{figure}

The discrepancy between theory and experiment can be further
understood by looking at the $L_1,\,L_2,$~and~$L_3$ ionization cross
sections extracted from our experimental data. The ionization cross
sections obtained from our experiment, along with theoretical
estimates, are plotted in the Figs.~\ref{Fig:Pb_Ionization} and
~\ref{Fig:Th_Ionization} for Lead and Thorium respectively. In both
the elements, the $L_3$ ionization cross section is explained very
well by the DWBA theory, specifically for the energies $E>1.35U$,
where $U$ is the ionization threshold for the $L_3$ subshell. Also it
is important to note that the $L_3$ subshell ionization cross sections
for only a handful of elements in the range from Phosphorus($Z=15$) to
Uranium ($Z=92$), measured either directly from electron energy loss
spectroscopy (EELS) or indirectly by electron impact spanning energy
range from near the ionization threshold to $\sim 1$~MeV, are found
to agree reasonably well with the DWBA calculations following Bote et
al.\cite{bote2008}, as described in detail in Ref.~\cite {Llovet2014}.
The agreement is very limited especially at energies near the
ionization threshold. But in case of $L_2$ and $L_1$ subshells, the
agreement between theory and experiment is not at all satisfactory. In
both the elements studied in our experiment, the $L_2$ and $L_1$
subshell ionization cross sections at near-threshold energies are
smaller than the DWBA estimates by $\sim 30 - 50\%$. The MRBEB theory
predicts $\sim 20 - 30\%$ higher ionization cross sections than the
DWBA theory for $L_2$ and $L_3$ subshells and up to $80 \%$
higher for
$L_1$ subshell.

The difference in theory and experiment for $L_1$ and $L_2$ sub-shell
results can be due to the relaxation parameters used in the estimation
of these ionization cross sections. A direct indication of the results
of relaxation parameter variation is given in Sec.~\ref{sc:data} in
connection with our attempt in extracting $\sigma_{L_1}$ and
$\sigma_{L_2}$ for Thorium from the corresponding $\sigma_{L_\alpha}$
and $\sigma_{L_\gamma}$. While calculating the $\sigma_{L_1}$ from the
corresponding $\sigma_{L_{\gamma_{236}}}$ for both the elements, we
have used the radiative yields $\Gamma_{\gamma_1},\Gamma_{\gamma_6},
\Gamma_{\gamma_2},\Gamma_{\gamma_3}$ and the fluorescence yield
($\omega_1$). Out of these five relaxation parameters,
$\Gamma_{\gamma_2},\Gamma_ {\gamma_3}$ and $\omega_1$ are associated
with the relaxation of the vacancy created in the $L_1$ subshell, and
the remaining parameters are associated with the vacancy in the $L_2$
subshell. Thus, our experimental findings indicate that the
differences between theory and experiment could be due to the poorly
known relaxation parameters, specifically the relaxation parameters
related to the $L_1$ subshell. It is worth mentioning here that in a
review on theories of inner shell ionization by proton
impact\cite{Miranda2002}, the author has concluded that the radiative
yield related to the $L_1$ subshell and the Coster- Kronig factors
need to be re-evaluated and experimentally measured.

The $L_2$ subshell results are inconclusive due to the fact that the
$\sigma_{L_2}$, obtained from $\sigma_{L_{\gamma_{1+5}}}$ is lower
than the theoretical estimates by ~$30-50\%$, but the
$\sigma_{L_\beta}$ values, which have almost equal contribution from
$\sigma_{L_2}$ and $\sigma_{L_3}$, are explained reasonably well by
the DWBA theory, specifically for Lead and other high $Z$
elements\cite{Varea2011,Rahangdale2014}. The $\sigma_{L_3}$ and
$\sigma_{L_\alpha}$ results are explained very well by theory, not
only for the Pb and Th, but also for the other high $Z$
elements\cite{Varea2011,Rahangdale2014}, indicating that the
relaxation parameters related to $L_3$ subshell are consistent with
the underlying theory and related experiments.

From our study, it is evident that the discrepancy between theory and
experiment may arise due to errors in fixing some of the relaxation
parameters. It is, therefore, important to perform measurements, which
require a minimum number of relaxation parameters for extracting
ionization cross sections from the experimental data. Clearly, more
measurements with wavelength dispersive spectrometer should be done
where resolution is so high that even a single transition can be
studied, thereby reducing the dependence on the relaxation parameters.
Also, very few measurements exist for the $L_\gamma$ X-ray production
cross sections of high $Z$ elements. As $L_\gamma$ transitions relate
to the $L_1$ and $L_2$ subshells, it is  important to perform these
measurements, specifically in view of the new calculations performed
by Pindzola ~\cite{Pindzola2014,Pindzola2015} by the inclusion of the
retarded electromagnetic potential, which significantly changes the
ionization cross sections of the $L_1$ and $L_2$ subshells.

\section{Conclusion}
\label{sc:con}

We have obtained the subshell resolved ionization cross sections from
the production cross sections involving the $L$-shell in Lead and
Thorium. Results are compared with two different theoretical
formalisms {\em viz.}, MRBEB and DWBA. The experimental results are
reproduced reasonably well by the DWBA theory for the $L_3$ subshell
and the $L_\alpha$ transition, but poor agreement is found for $L_1$
and $L_2$ subshells and consequently for the $L_\beta$ and the
$L_\gamma$ transitions in both the elements. MRBEB theory overpredicts
the cross sections for all the three subshells in both the elements in
the electron impact energy regime explored in our experiment.
Discrepancy between DWBA theory and our experiment points to the poor
knowledge of the relaxation parameters related to the $L_1$ subshell.
From our study, we conclude that more precise measurements of the
corresponding sub-shell resolved cross sections are urgently needed to
obtain the relaxation parameters with better precision.

%================ACKNOWLEDGMENTS======================== %
%================ACKNOWLEDGMENTS======================== %

\begin{acknowledgments}
This research work is supported by grant no. 12-R\&D-SIN-5.02-0102 of
the Department of Atomic Energy, Government of India. Two of the
authors (M.G. and J.P.S.) would like to acknowledge research support
in part by Funda\c{c}\~{a}o Fundacao para a Ci\^{e}ncia eTecnologia
(FCT), Portugal, through the Project No. PEstOE/FIS/UI0303/2011
financed by the European Community Fund FEDER through the COMPETE. M.
G. also acknowledges the support of  the FCT, under Contract No.
SFRH/BPD/92455/2013. The authors acknowledge the fabrication of the
experimental setup by the SINP machine shop.
\end{acknowledgments}

\altaffiliation{ Laborat\'{o}rio de Instrumenta\c{c}\~{a}o, Engenharia
Biom\'{e}dica eF\'{\i}sica da Radia\c{c}\~{a}o (LIBPhys-UNL), Departamento de F\'{\i}sica, Faculdade de
Ci\^{e}ncias e Tecnologia, FCT, Universidade Nova de Lisboa, 2829-516
Caparica, Portugal}

\bibliographystyle{apsrev4-1}

\end{document}